\documentclass[pdflatex,sn-mathphys]{sn-jnl}

\jyear{2022}%

\theoremstyle{thmstyleone}%
\theoremstyle{thmstyletwo}%

\theoremstyle{thmstylethree}%

\usepackage{subfigure}
\usepackage{comment}
\usepackage{arydshln} 

\begin{document}

\title[The double-LGAD]{A new Low Gain Avalanche Diode concept: the double-LGAD}

\author*[1]{\fnm{F.} \sur{Carnesecchi}}\email{francesca.carnesecchi@cern.ch}
\author*[2,5]{\fnm{S.} \sur{Strazzi}}\email{sofia.strazzi2@unibo.it}
\author[2,5]{\fnm{A.} \sur{Alici}}
\author[3,4]{\fnm{R.} \sur{Arcidiacono}}
\author[3]{\fnm{N.} \sur{Cartiglia}}
\author[5]{\fnm{D.} \sur{Cavazza}}
\author[7]{\fnm{S.} \sur{Durando}}
\author[3]{\fnm{M.} \sur{Ferrero}}
\author[5]{\fnm{A.} \sur{Margotti}}
\author[3,6]{\fnm{L.} \sur{Menzio}}
\author[5]{\fnm{R.} \sur{Nania}}
\author[2,5]{\fnm{B.} \sur{Sabiu}}
\author[2,5]{\fnm{G.} \sur{Scioli}}
\author[3]{\fnm{F.} \sur{Siviero}}
\author[3,6]{\fnm{V.} \sur{Sola}}
\author[5]{\fnm{G.} \sur{Vignola}}

\affil[1]{\orgname{CERN}, \orgaddress{\city{Geneva}, \country{Switzerland}}}
\affil[2]{\orgdiv{Dipartimento Fisica e Astronomia dell’Università}, \orgaddress{\city{Bologna}, \country{Italy}}}
\affil[3]{\orgname{INFN}, \orgaddress{\city{Torino}, \country{Italy}}}
\affil[4]{\orgdiv{Università del Piemonte Orientale}, \orgaddress{\city{Novara}, \country{Italy}}}
\affil[5]{\orgname{INFN}, \orgaddress{\city{Bologna}, \country{Italy}}}
\affil[6]{\orgdiv{Università degli Studi di Torino}, \orgaddress{\city{Torino}, \country{Italy}}}
\affil[7]{\orgdiv{Dipartimento di elettronica e telecomunicazioni}, \orgname{Politecnico di Torino}, \city{Torino}, \country{Italy}}


\abstract{This paper describes the new concept of the double-LGAD. The goal is to increase the charge at the input of the electronics, keeping a time resolution equal or better than a standard (single) LGAD; this has been realized by adding the charges of two coupled LGADs while still using a single front-end electronics. 
The study here reported has been done starting from single LGAD with a thickness of  25\,\textmu{m}, 35\,\textmu{m} and 50\,\textmu{m}.}




\maketitle

\section{Introduction}\label{sec:intro}
Low Gain Avalanche Detectors (LGADs)\cite{2014PELLEGRINI}, also known as Ultra Fast Silicon Detectors (UFSDs) \cite{2017Sadrozinski}, are $n$-on-$p$ diodes with an additional highly doped $p^+$-type layer (gain layer) underneath the $n$-contact, which is responsible for the charge multiplication mechanism (in reverse bias regime).
It has been proven that LGADs with a thickness of 35\,\textmu{m} combined with a gain G$\sim30$ can provide a time resolution around 22\,ps \cite{2023Carnesecchi}. \\
Thanks to the excellent timing performance, this technology is already envisioned or proposed for several detector upgrades and applications \cite{atlas,cms,alice3}.

It has already been demonstrated that the time resolution of LGADs improves with thinner designs. Nevertheless, a thinner design also implies a smaller charge at the input of the amplifier which, because of the worse S/N, might worsen the performance of the amplifier and/or its power consumption. This need gave rise to the idea of the double-LGAD.

\section{The double-LGAD} 
\label{sec:d-LGAD}
\begin{figure}[h!!]
        \centering%
          {\includegraphics [width=0.6\textwidth]{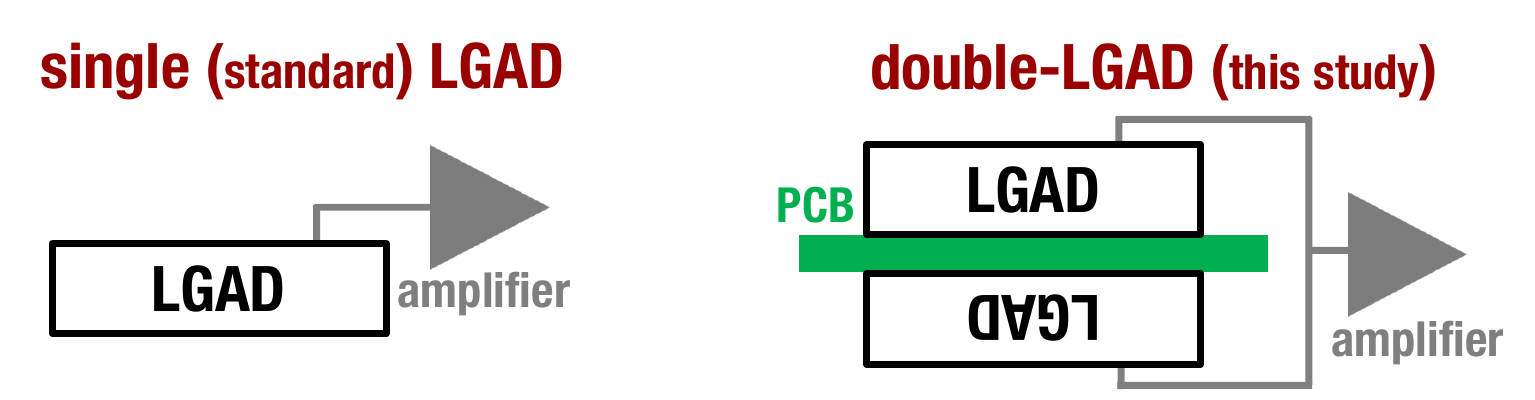}}
        \caption{Schematic of the single (standard) and double LGAD concept}
          \label{fig:concept}
\end{figure}

The concept of the double-LGAD (d-LGAD) is inspired by the Multigap Resistive Plate Chambers (MRPC) \cite{MRPC}. Essentially, the idea is to sum up the signal taken from a double-layer of LGADs, still using an unique front-end amplifier.\\ 
As a first step, the signals from two different LGADs have been summed using a specific PCB design (more details in Sec.\ref{sec:set_el}) and the output has been sent  to a single and
common amplifier.
In Figure \ref{fig:concept} a schematic of a d-LGAD is reported. This is currently just a proof of concept, but the natural next step would be a better integration of such a concept either in the board containing the electronics or in the detector itself (in a truly d-LGAD or e.g. using Through-Silicon Via, TSV, technique). \\
In the proposed scheme, given by the sum of two LGADs each with a certain thickness t, the charge of the d-LGAD, is expected to be double if compared with a single LGAD of same thickness t. 
The time resolution of such a d-LGAD is expected to be largely better than that of an equivalent single LGAD of thickness 2t. Similarly to MRPC \cite{MRPC_sim}, the time resolution of a d-LGAD is foreseen to improve also w.r.t. to a single LGAD of thickness t; however, due to different signal amplitudes in the two d-LGADs,  this improvement is not by a factor of $\sqrt{2}$.  As stated in \cite{MRPC_sim},  the time resolution will be dominated by the d-LGAD with the largest signal. In other words, in d-LGAD, the LGAD with the largest signal always dominates the time resolution. 


\section{Experimental setup} 
\label{sec:setup}

\subsection{Detectors} 
\label{sec:det}

\label{sec:detectors}
The tested LGADs came from two manufacturers, Hamamatsu Photonics K.K. (HPK, Japan) and Fondazione Bruno Kessler (FBK, Italy) and have a different area (A). The sensors from HPK have a nominal thickness (T)\footnote{Usually the active thickness is around 2-3 \textmu{m} less than the nominal. } of 50 \textmu{m}, appertaining to a very uniform wafer.
The FBK LGADs\footnote{This UFSD production is called EXFLU0~\cite{exflu}.} have a nominal thickness of 25 \textmu{m} and 35 \textmu{m}, respectively. More details about the FBK LGADs characteristics and performances can be found in \cite{2023Carnesecchi}. \\
\begin{table}[ht]
\begin{center}
\begin{minipage}{\textwidth}
\caption{Characteristics of the Front (F) and Back (B) LGADs of each couple under test.} \label{tab:lgad_char}
\begin{tabular*}{\textwidth}{@{\extracolsep{\fill}}lcccccc@{\extracolsep{\fill}}}
\toprule
 & A (mm$^2$)  & T (\textmu{m}) & V$_{bd}$ (V) & V$_\text{applied} (V)$  & Gain\\
\midrule
FBK25-F & 1 $\times$ 1 & 25 & 132 $\pm$ 1  & 80-120   & 11-24\\
FBK25-B & 1 $\times$ 1 & 25 & 124 $\pm$ 1 & 80-120   & 12-43\\
FBK35-F & 1 $\times$ 1  & 35  & 266.5 $\pm$ 0.5  & 180-240   & 9-17\\
FBK35-B & 1 $\times$ 1  & 35 & 268 $\pm$ 1  & 190-240   & 11-27\\
HPK50-F & 1.3 $\times$ 1.3 & 50  & 224.6  $\pm$ 0.2 &  170-220 & 24-63\\
HPK50-B & 1.3 $\times$ 1.3 & 50 & 237.4  $\pm$ 0.2 &  170-220 & 25-64\\
\botrule
\end{tabular*}
\end{minipage}
\end{center}
\end{table}
All the tested detectors have been previously completely characterized at the INFN Bologna laboratories. The method to measure the breakdown voltage (V$_{bd}$) gain are explained in \cite{2023Carnesecchi}.
The main characteristics of the sensors are reported in Table \ref{tab:lgad_char}.

\subsection{Beam test setup and electronics}
\label{sec:set_el}
The time resolution of the UFSDs has been studied at the T10 beamline at PS-CERN in July and November 2022. The beam was mainly composed of protons and pions with a momentum of +10 GeV/c.
For each data acquisition up to 4 carrier boards mounted on micro-mover stages were aligned to the beam in a telescope frame at a relative distance of 24 cm, and the whole setup was enclosed in a dark environment box at room temperature.

All the LGADs tested have been mounted on a board V1.4-SCIPP-08/18, containing a wide bandwidth (2 GHz) and low noise inverting amplifier with a measured amplification of factor 6. The board has been modified in order to place one LGAD on each side of the board and, thanks to a TSV in the PCB itself, connect the output of the two LGADs together and send the signal to the amplifier described above, realizing a first prototype of d-LGAD.
The output of the board was followed by a second amplification stage, with a gain factor of around 13 and 14 respectively for the HPK50 and FBK sensors \footnote{The second amplifier used for the HPK50 and for the FBK sensors were the minicircuit LEE39+ (\href{https://www.minicircuits.com/pdfs/LEE-39+.pdf}{LEE39+ datasheet}) and Gali52+ (\href{https://www.minicircuits.com/pdfs/GALI-52+.pdf}{GALI52+ datasheet}) respectively.}.

Up to 4 amplified signals were sent to a LeCroy WaveRunner 9404M-MS oscilloscope\footnote{\href{https://teledynelecroy.com/oscilloscope/waverunner-9000-oscilloscopes/waverunner-9404m-ms
}{Lecroy WaveRunner datasheet}}, with 20 Gs/s sampling rate,  4 GHz of analogue bandwidth and 8-bit vertical resolution. The contribution of the oscilloscope time resolution to the measured one was negligible.

For the trigger of the data acquisition, a threshold has been set for each channel and the coincidence of the four sensors has been used.

\begin{figure}[htbp]
    \centering
    \subfigure[\label{fig:rms_sn_25}]{\includegraphics[width=0.48\textwidth]{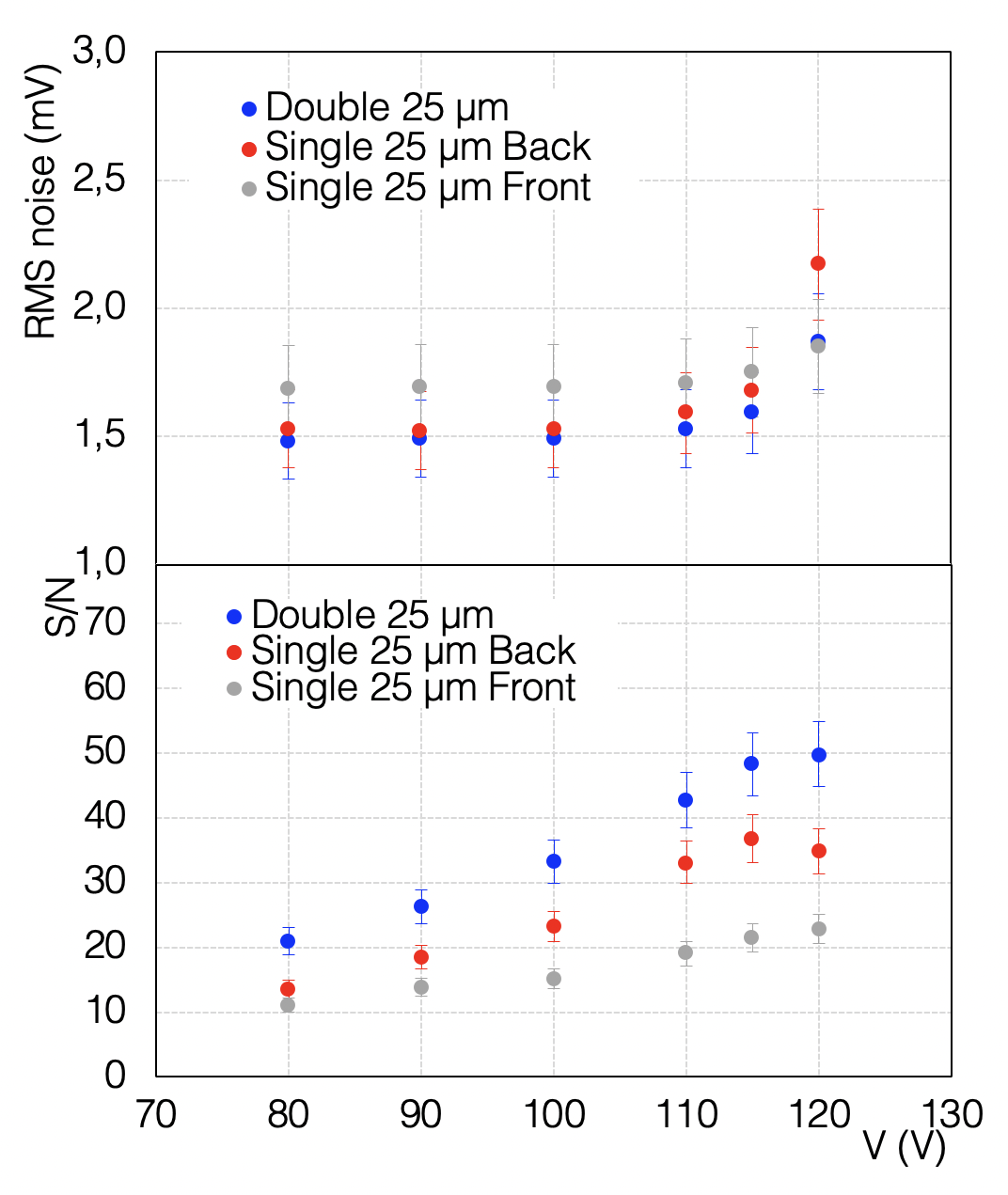}} 
    \subfigure[\label{fig:rms_sn_35}]{\includegraphics[width=0.48\textwidth]{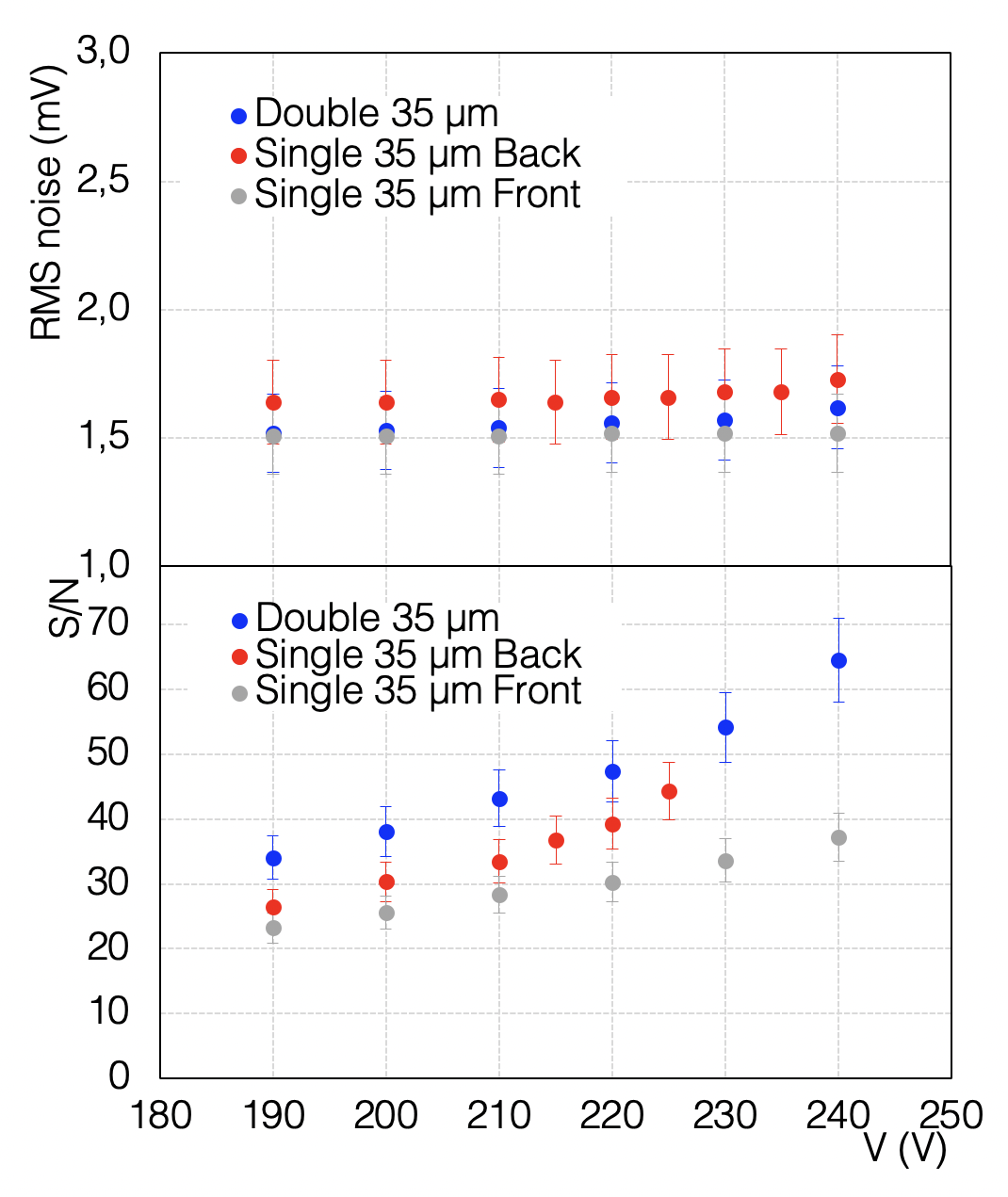}} 
    \subfigure[\label{fig:rms_sn_50}]{\includegraphics[width=0.48\textwidth]{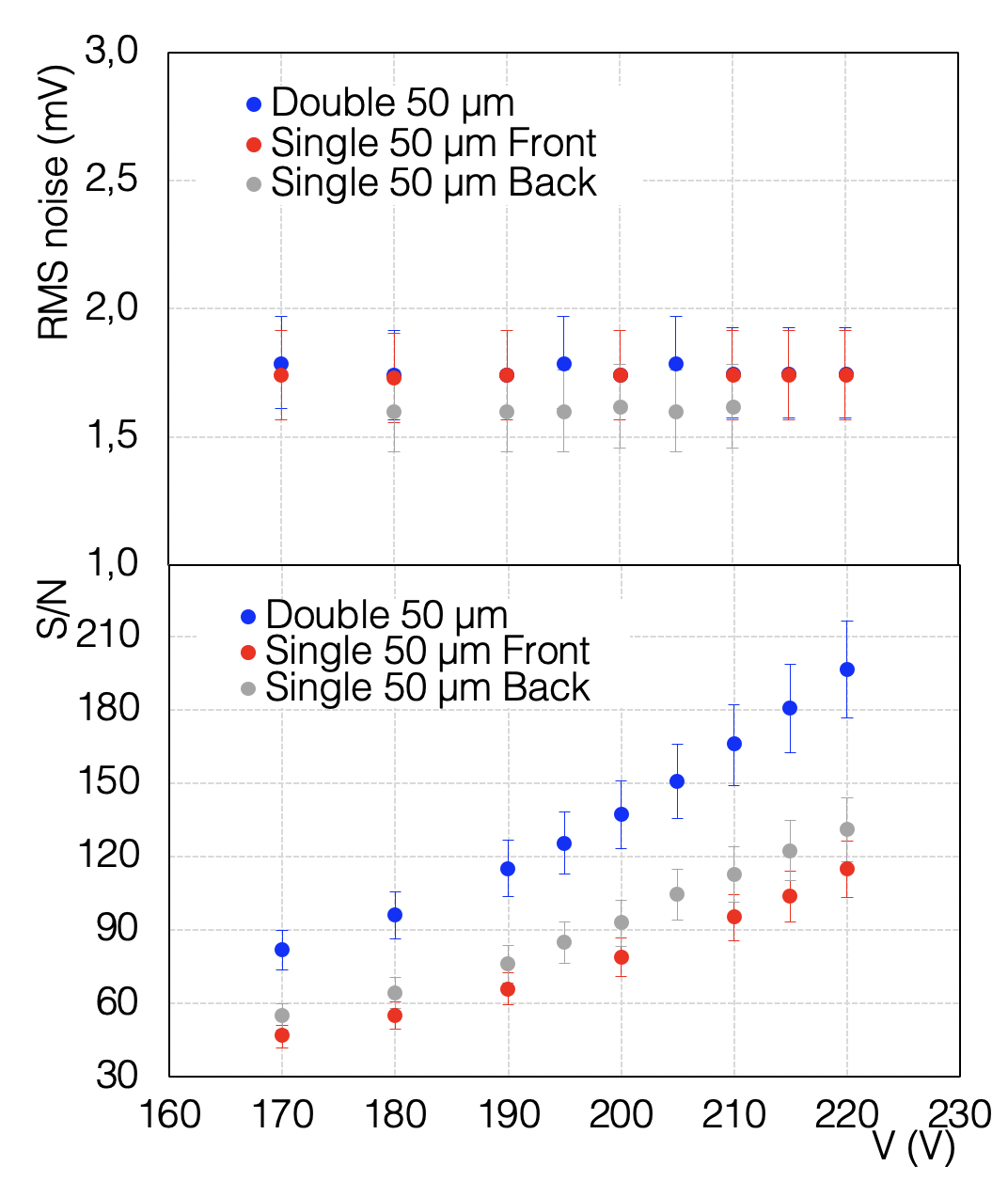}}
    \caption{RMS of the noise and S/N ratio for all the tested LGADs.}
    \label{fig:rms}
\end{figure}

For all the measurements reported in this paper, the double LGAD has always been compared with the performances of the single LGADs composing the d-LGAD under test\footnote{First we tested always the d-LGAD. Then we un-bonded the bottom LGAD in order to test the top one. Lastly, we un-bonded the top LGAD, bonding and testing the bottom one}. Therefore, every LGAD thickness will always have three different measurements: two coming from each of the LGAD composing the d-LGAD, and one from the d-LGAD itself.

The RMS of the noise (see \cite{2023Carnesecchi} for more details) and the Signal to Noise ratio (S/N) have been evaluated for each sensor and voltage. In Figure \ref{fig:rms} they are reported as a function of the applied voltage.
As can be seen, the noise between single and double sensors is compatible for all thicknesses. The S/N instead is always higher for the d-LGAD, giving already some insight into the better performances reported later in the paper.\\

\section{Results} 
\label{sec:results}
The data analysis was performed following similar procedures to those reported in \cite{2019Carne, 2023Carnesecchi}. In particular, thanks to the oscilloscope readout, the full signal waveforms were recorded and analyzed. It was then possible to use the Constant Fraction Discriminator (CFD) method to extract the DUTs time resolutions.
Moreover, to filter out the high-frequency noise, a smoothing of the LGAD signal was applied with a four-point moving average.

To extract the time resolution of a single DUT (single or double-LGAD), a system with  three sensors and three differences between the arrival time of each pair of detectors has been considered. The sigma extracted from the fit has then been used to obtain the final time resolution of the three LGADs at a given voltage and CFD.\\

\begin{figure}[htbp]
        \centering%
        \subfigure[\label{fig:charge_25}]%
          {\includegraphics [width=0.45\textwidth] {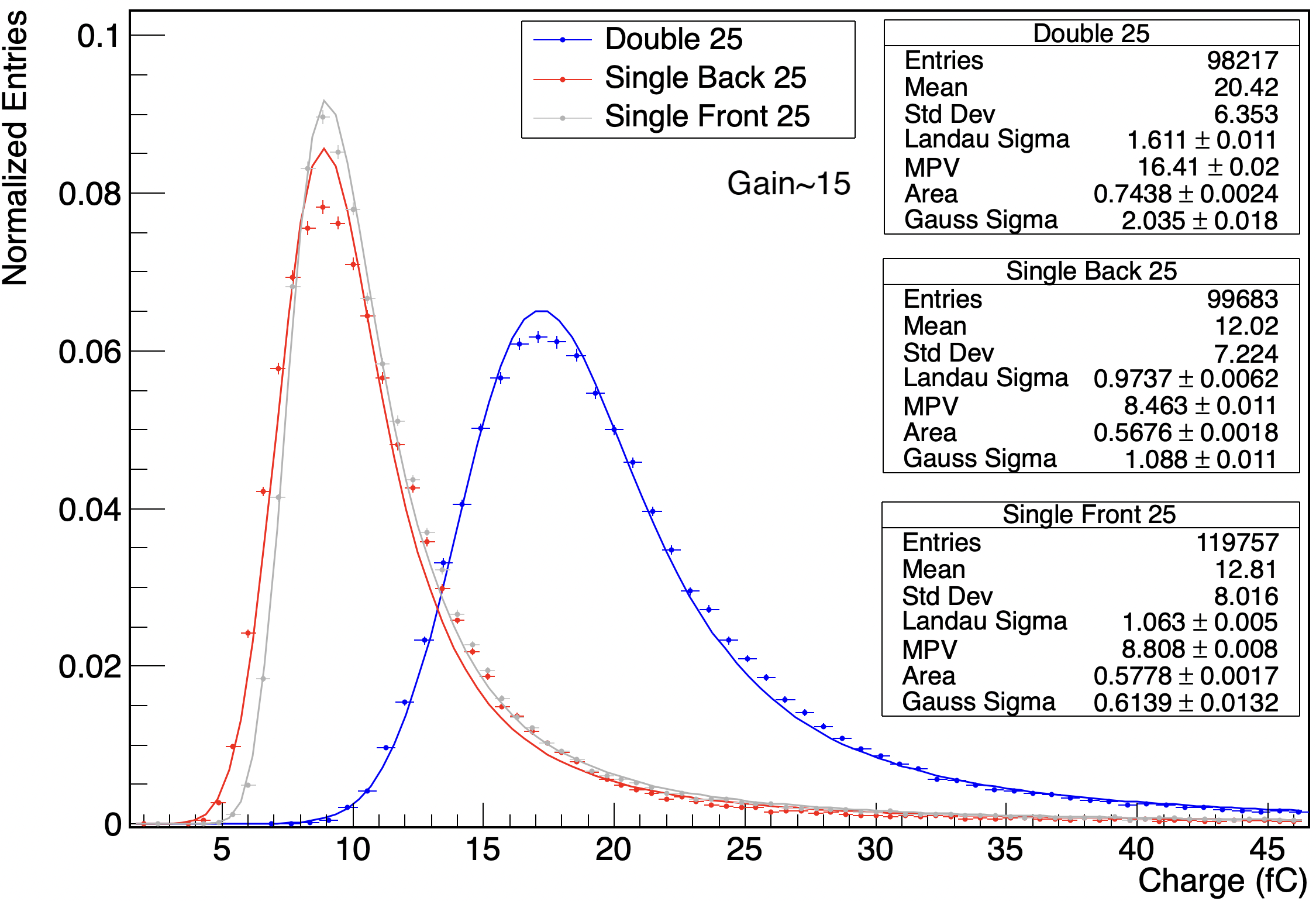}}\quad
        \centering%
        \subfigure[\label{fig:charge_35}]%
          {\raisebox{-0.0cm}{
          	\includegraphics[width=0.45\textwidth]{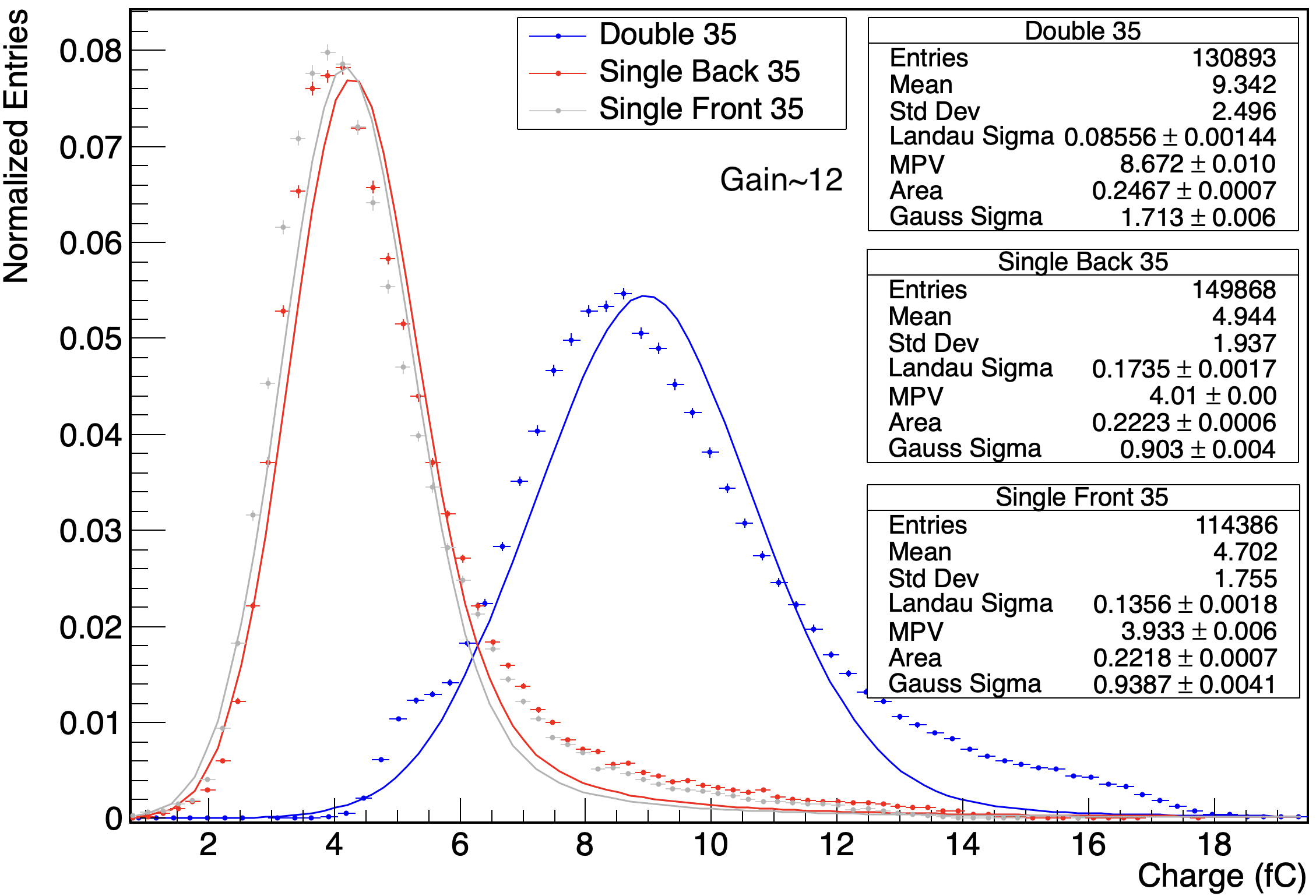}}}
        \subfigure[\label{fig:charge_50}]%
          {\raisebox{-0.0cm}{
          	\includegraphics[width=0.45\textwidth]{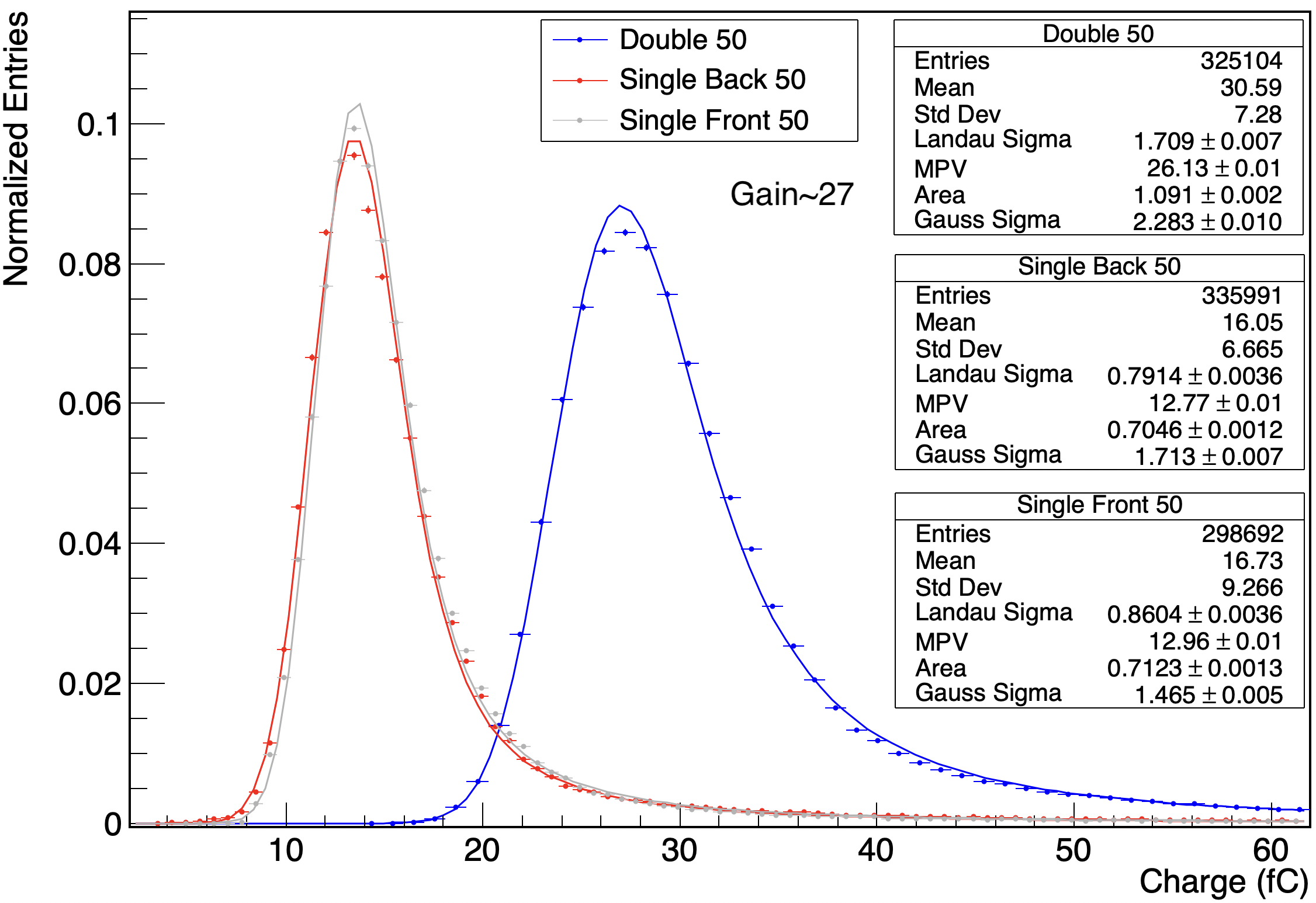}}}
          	\centering%
        \caption{(a) Charge distributions for all the DUTs at a gain of 15, 12 and 27 for the FBK25, FBK35 and  HPK50 couples. The distributions are fitted with a convolution of a Gaussian and a Landau functions.}
        \label{fig:landau}
\end{figure}

In Figure \ref{fig:landau} the measured charge distributions are shown. Notice that, the comparisons between single and the double LGADs has been done at a fixed gain.
As can be observed the d-LGAD always showed an MPV of charge which was double that of the single one, as naively expected, demonstrating the success of a so-built detector and, as a consequence, the potentially less demanding electronic front-end that can be realized, thanks to the larger charge integrated in input.\\

\begin{figure}
        \centering%
          \subfigure[\label{fig:time_v_25}]%
          {\includegraphics [width=5cm] {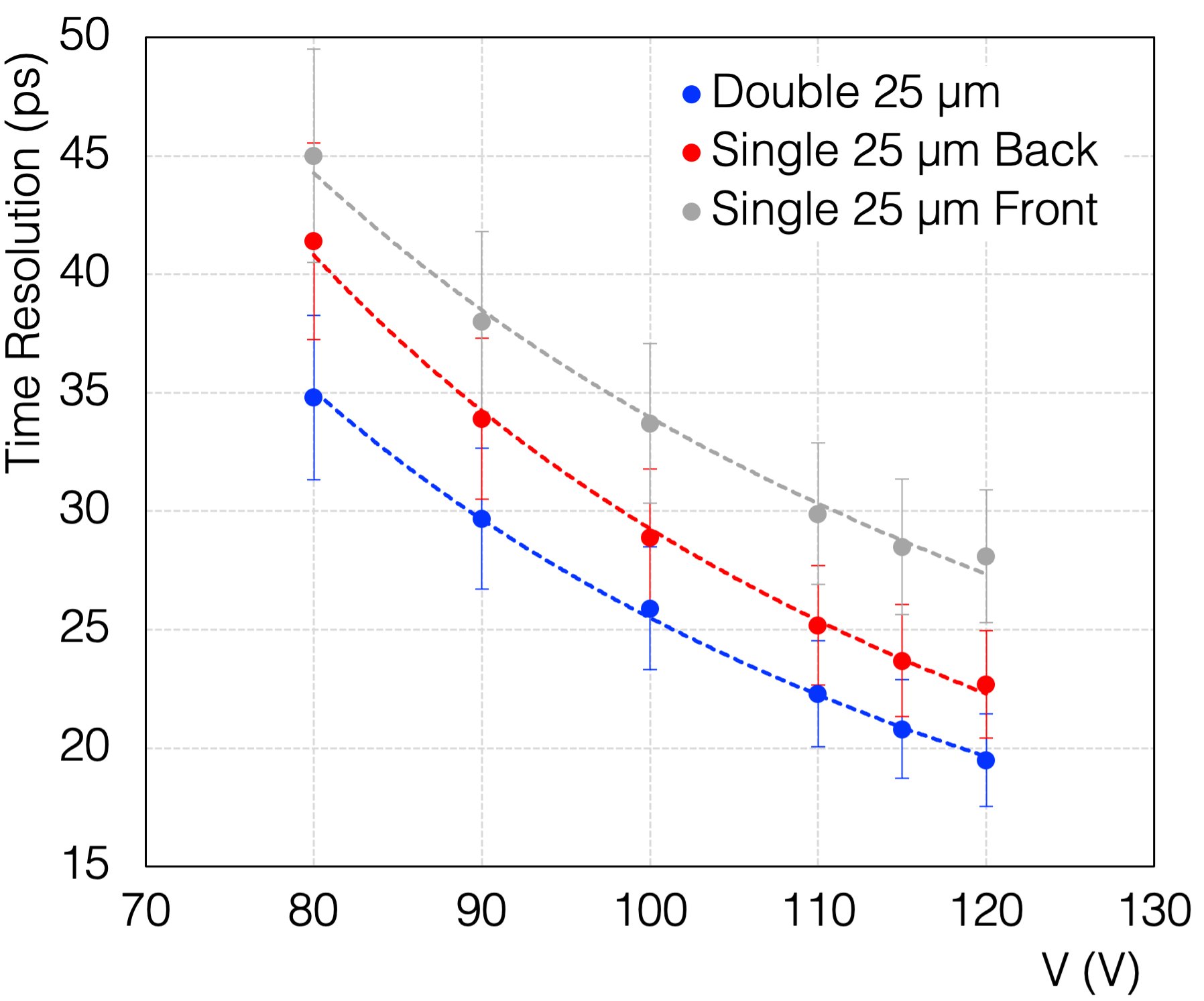}}
        \centering%
        \subfigure[\label{fig:time_v_35}]%
          {\raisebox{-0.0cm}{
          \includegraphics[width=5cm]{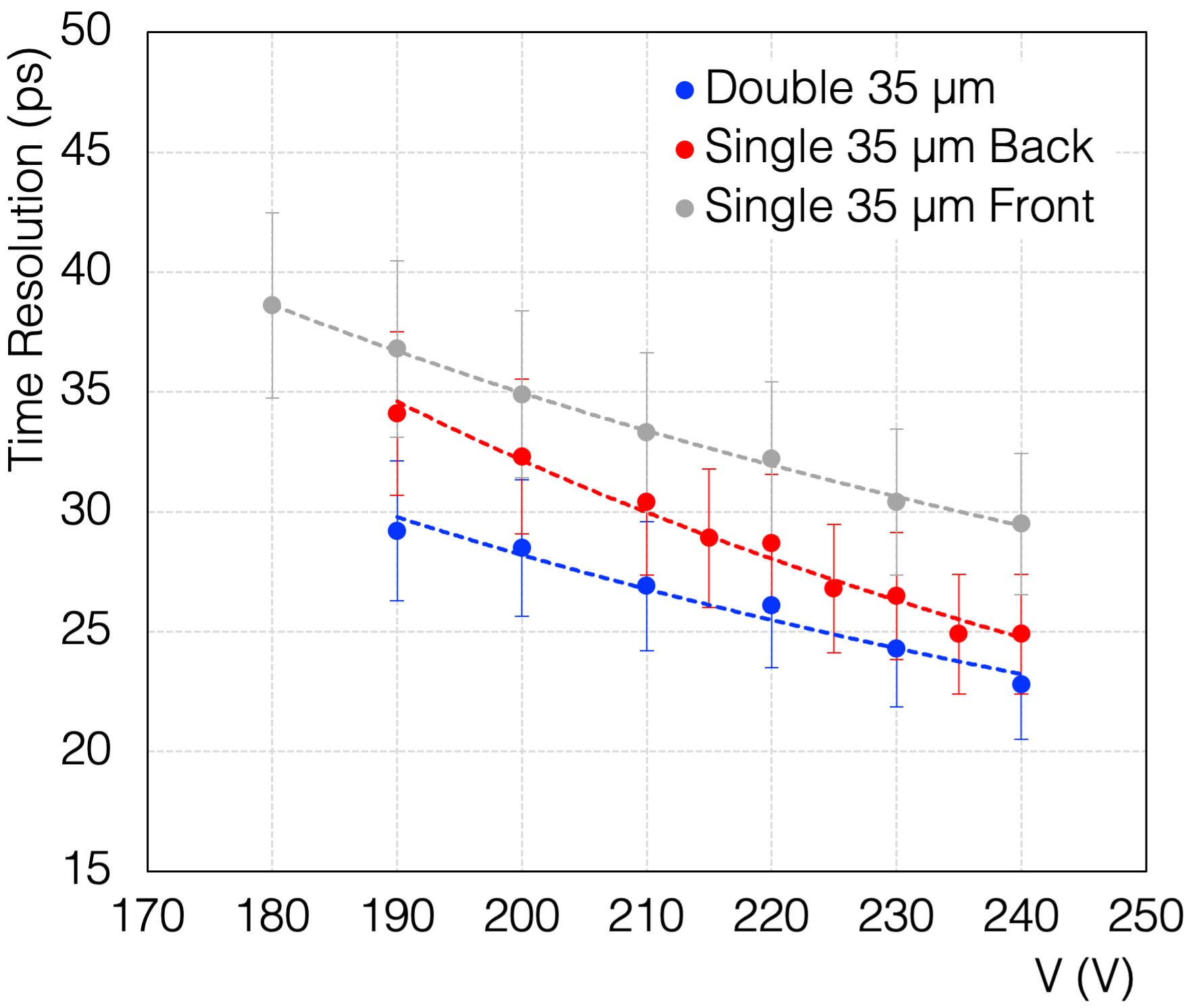}}}
        \centering%
          \subfigure[\label{fig:time_v_50}]%
          {\includegraphics [width=5cm] {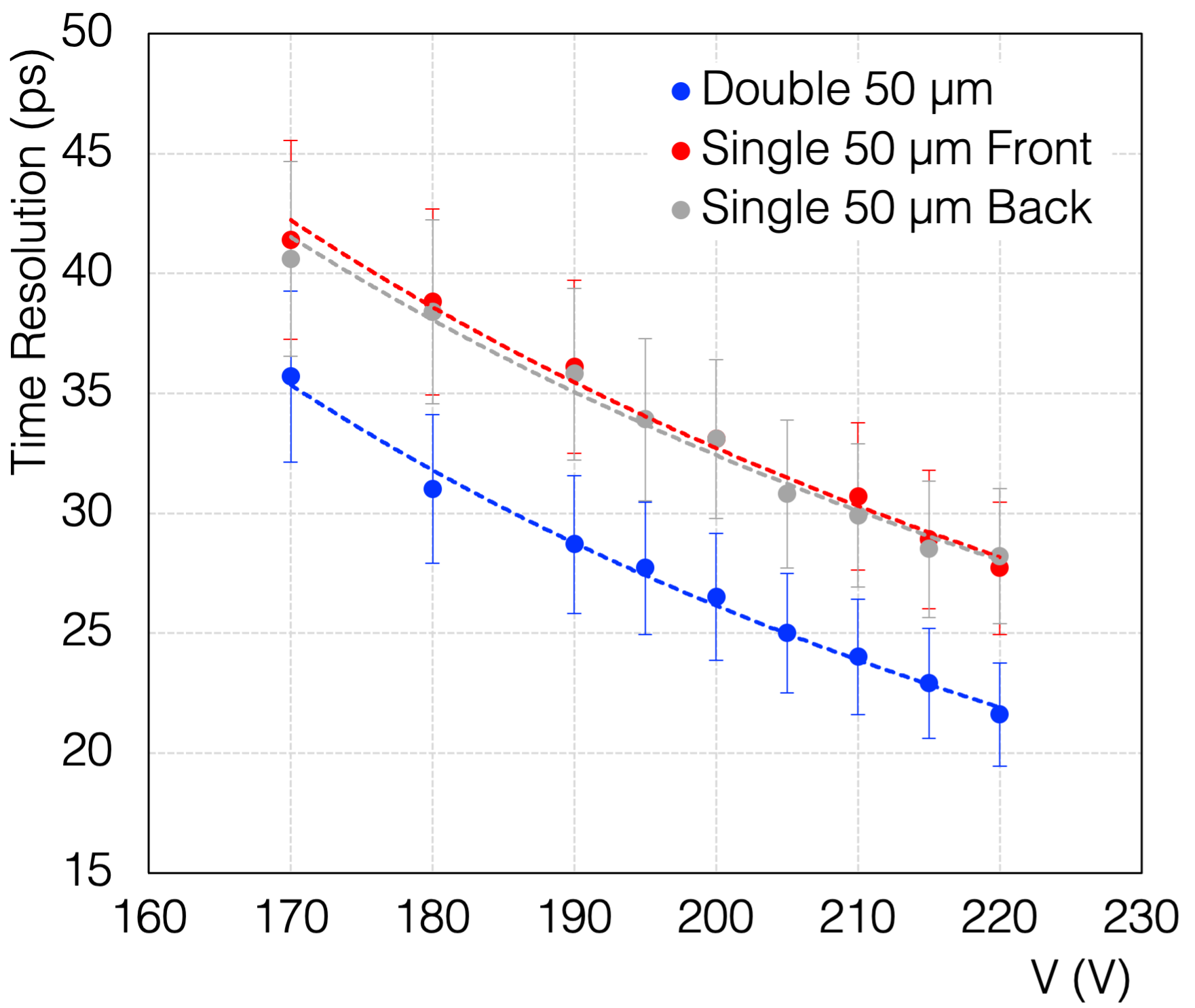}}
          	\centering%
        \caption{\label{fig:time_v} Measured time resolution results from the beam test as a function of the voltage applied for all the LGADs tested 25, 35 and 50 for a CFD of 50\%, 30\% and 30\%, respectively. The errors for the measured time resolution have been estimated as 10$\%$ of the value.  The lines are included to guide the eye.}
\end{figure}

In all the following plots, the measured time resolution for a fixed CFD (more details in \cite{2019Carne, 2023Carnesecchi}) has always been considered.
In Figure \ref{fig:time_v} the time resolution as a function of the applied voltage is reported.

First of all, if compared with \cite{2023Carnesecchi}, the results for all thicknesses are compatible.
It can be noticed that the resolutions of the single LGADs are very nicely uniform only for the 50 \textmu m couple, owing to the more uniform sensor wafer (and specifically to the more similar gains for the two sensors).
Nevertheless for all three thicknesses, for a fixed voltage an improvement of the time resolution has been observed with the d-LGADs. A final time resolution of $\sim$~20~ps has been obtained for all three thicknesses. 
For the future, new 25 and 35 \textmu m  production with increased uniformity could potentially improve the time resolution of d-LGAD, bringing time resolution below 20 ps.

\begin{figure}
        \centering%
          \subfigure[\label{fig:time_charge_25}]%
          {\includegraphics [width=5cm] {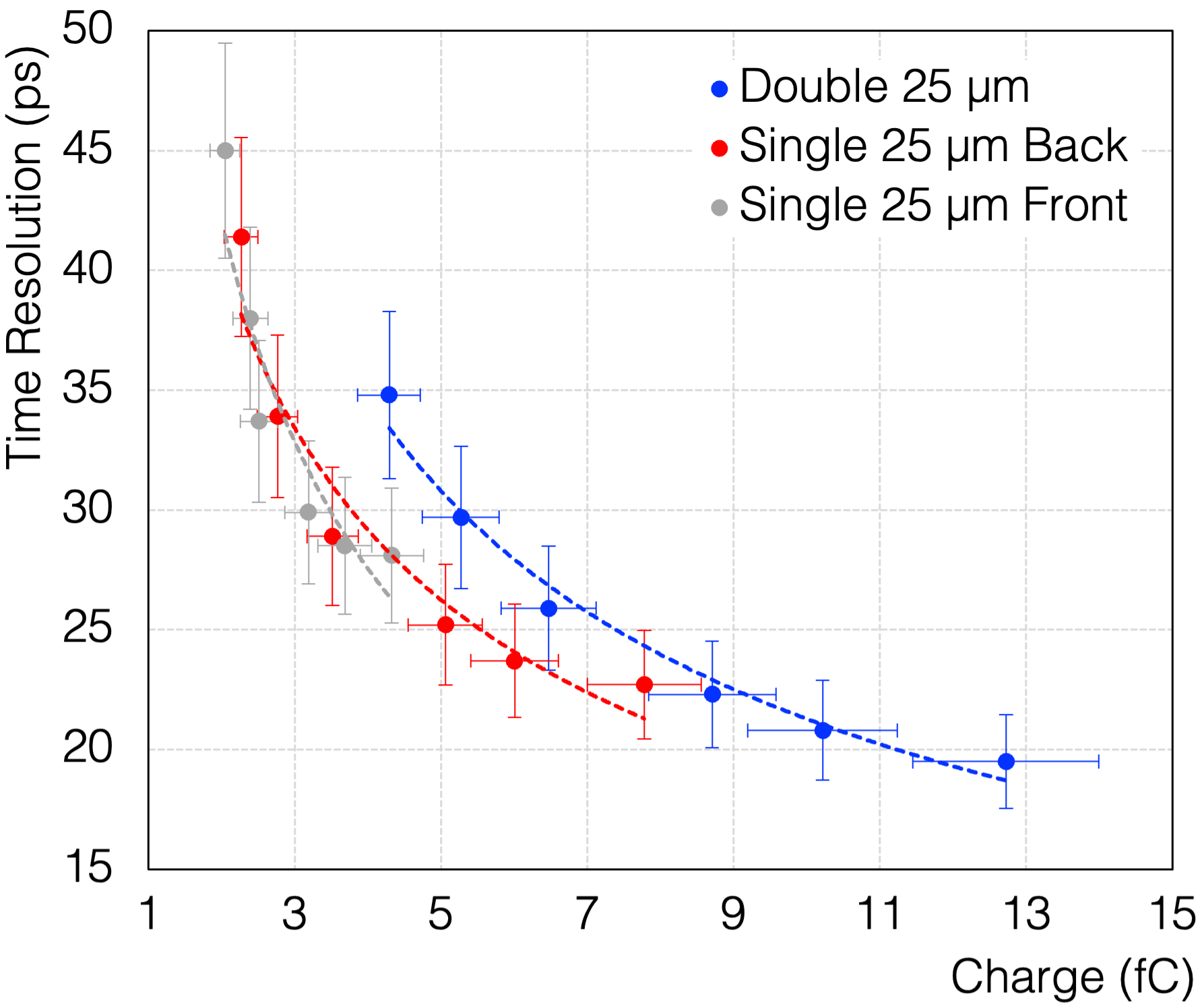}}
        \centering%
        \subfigure[\label{fig:time_charge_35}]%
          {\raisebox{-0.0cm}{
          	\includegraphics[width=5cm]{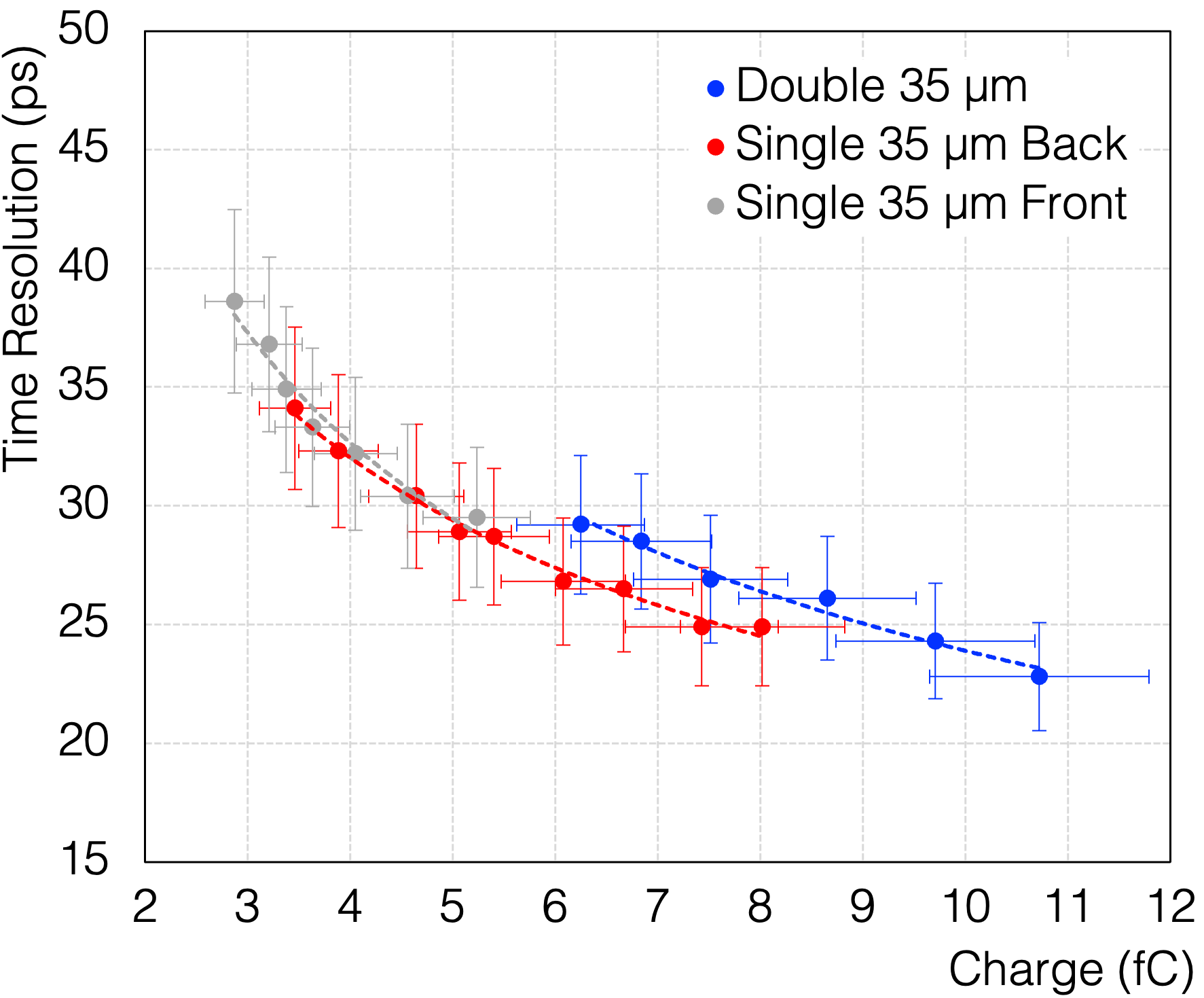}}}
        \centering%
          \subfigure[\label{fig:time_charge_50}]%
          {\includegraphics [width=5cm] {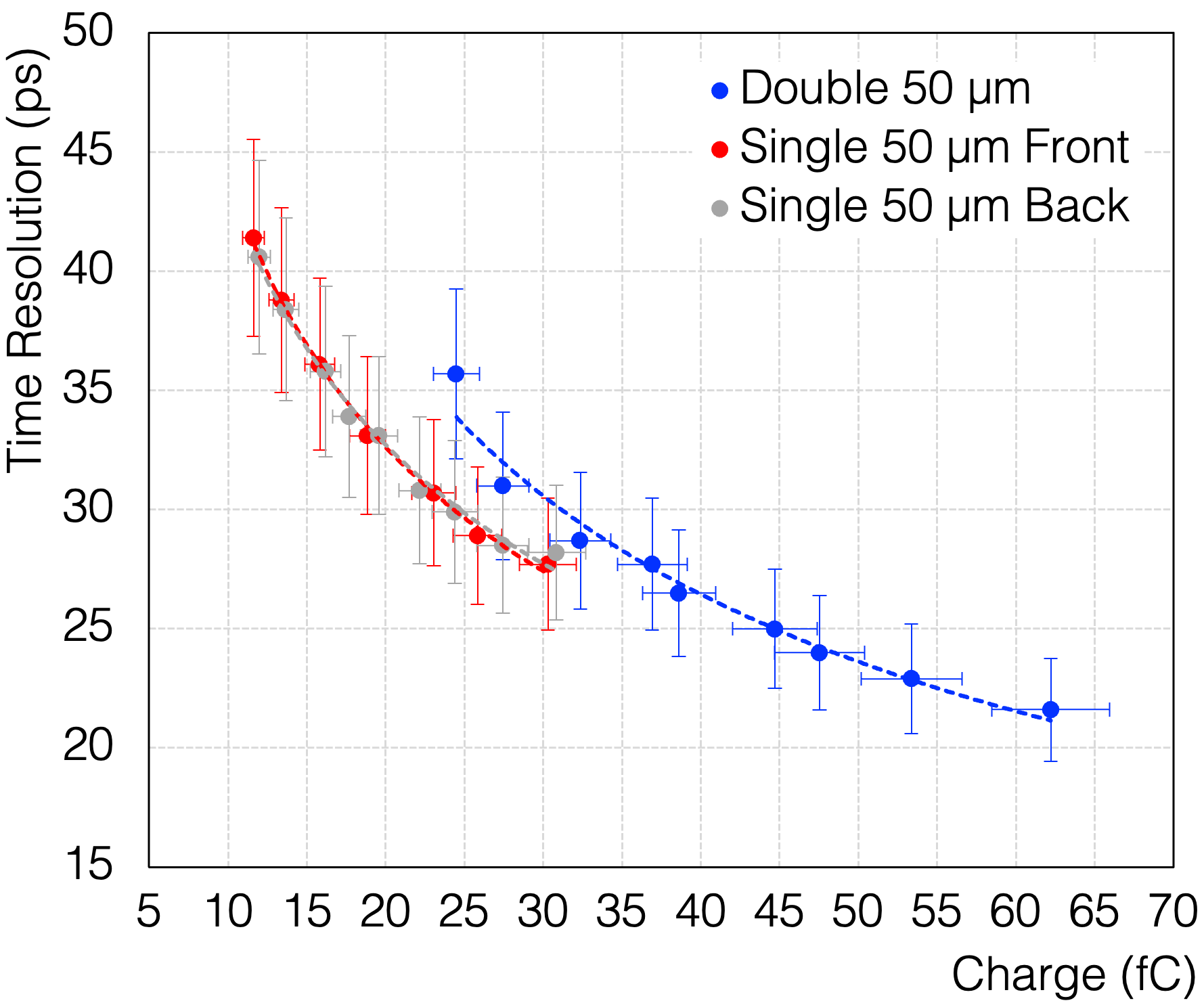}}
          	\centering%
        \caption{\label{fig:time_charge_gain} Measured time resolution results from the beam test as a function of the charge collected for all the LGADs tested, 25, 35 and 50 \textmu m for a CFD of 50\%, 30\% and 30\%, respectively. The errors for the measured time resolution have been estimated as 10$\%$ of the value. The lines are included to guide the eye.}
\end{figure}

In Figure \ref{fig:time_charge_gain} the time resolution as a function of the charge is then reported. The plot would be totally similar if plotted vs gain. 
As expected, the d-LGADs show higher charge wrt to single LGADs at the input of electronics, in particular for uniform wafers and couples, as in the case of the 50 \textmu m  thickness.

\begin{table}[ht]
\begin{center}
\begin{minipage}{\textwidth}
\caption{Time resolution for 25, 35 and 50 for a given voltage (or gain) obtained in a beam test setup at room temperature.} \label{tab:res}
\begin{tabular*}{\textwidth}{@{\extracolsep{\fill}}lccc@{\extracolsep{\fill}}}
\toprule
 & Voltage applied & Gain & Time resolution\\
\midrule
FBK25 Front & 120 V & 24 $\pm$ 2 & (23 $\pm$ 2) ps\\
FBK25 Back & 120 V & 43 $\pm$ 4 & (28 $\pm$ 3) ps\\
d-FBK25 & 120 V & 35 $\pm$ 4 & (20 $\pm$ 2) ps\\
\cdashline{1-4}
FBK35 Front & 240 V & 17 $\pm$ 2 & (30 $\pm$ 3) ps\\
FBK35 Back & 240 V & 27 $\pm$ 3 & (25 $\pm$ 2) ps\\
d-FBK35 & 240 V & 15 $\pm$ 2 & (23 $\pm$ 2) ps\\
\cdashline{1-4}
HPK50 Front & 220 V & 63 $\pm$ 6 & (28 $\pm$ 3) ps\\
HPK50 Back & 220 V & 64 $\pm$ 6 & (28 $\pm$ 3) ps\\
d-HPK50 & 220 V & 59 $\pm$ 6 & (22 $\pm$ 2) ps\\

\botrule
\end{tabular*}
\end{minipage}
\end{center}
\end{table}

In Table \ref{tab:res} the best time resolutions reached for the three detectors tested are summarized.

\section{Conclusions}\label{sec:conclusions}
The study presented in this paper describes a new concept for improving time resolution by coupling two LGADs connected to the same amplifier. All the results have been obtained using a 10 GeV/c beam at CERN PS. 
For three different thicknesses of sensors, the performance of the d-LGAD has been compared with that of the single LGADs composing it. The d-LGAD concept shows clear advantages over the standard single LGAD, resulting in better time resolution with the added benefit of a higher charge provided at the input of the amplifier. In particular, results demonstrate a consistent improvement in time resolution for the d-LGAD compared to the single LGADs, reaching a time resolution of $\sim$~20~ps for all three thicknesses. 
Additionally, for all the couples, the charge MPV generated by the d-LGAD is doubled compared to both single sensors, as expected, resulting in a clear advantage for the electronics.
Overall, this concept presents a promising development for LGAD’s performance and paves the way for future implementation of such sensors.

\backmatter

\bmhead{Acknowledgments}

We acknowledge the following funding agencies and collaborations: INFN – FBK agreement on sensor production; Dipartimenti di Eccellenza, Univ. of Torino (ex L. 232/2016, art. 1, cc. 314, 337); Ministero della Ricerca, Italia, PRIN 2017, Grant 2017L2XKTJ – 4DinSiDe; Ministero della Ricerca, Italia, FARE,    Grant R165xr8frt\_fare.\\
The authors wish also to thank the support of the mechanical and electronic workshops of the INFN Unit of Bologna and the CERN-PS operator team for the support. We would also like to thank the CERN Bondlab for their availability during the beam tests.

\section*{Declarations}
The study was funded by: INFN – FBK agreement on sensor production; Dipartimenti di Eccellenza, Univ. of Torino (ex L. 232/2016, art. 1, cc. 314, 337); Ministero della Ricerca, Italia, PRIN 2017, Grant 2017L2XKTJ – 4DinSiDe; Ministero della Ricerca, Italia, FARE, Grant R165xr8frt\_fare.\\
The authors received research support from institutes as specified in the author list beneath the title. \\

\section*{Data availability}
The datasets generated during and/or analysed during the current study are available from the corresponding author on reasonable request.


\bibliography{sn-bibliography}


\end{document}